\documentclass[12pt]{iopart}
\usepackage{iopams}

\newcommand{\re}{\mbox{$\rm e$}}

\newcommand{\rd}{\mbox{$\rm d$}}

\begin{document}
\newtheorem{theorem}{Theorem}
\newtheorem{corollary}{Corollary}
\newtheorem{prop}{Proposition}

\title[Note on exponential families of distributions]
{Note on exponential families of distributions}

\author[D.~C.~Brody]{Dorje~C~Brody}

\address{Department of Mathematics, Imperial College,
London SW7 2BZ, UK}

\begin{abstract}
We show that an arbitrary probability distribution can be
represented in exponential form. In physical contexts, this implies
that the equilibrium distribution of any classical or quantum
dynamical system is expressible in grand canonical form.
\end{abstract}

\submitto{\JPA}
%\input{psfig.sty}
%
%  Uncomment out if preprint format required
%
%\pacs{00.00, 20.00, 42.10}
%\maketitle

Exponential families of probability distributions play central roles
in information theory~\cite{cover}, statistics~\cite{brown}, and
statistical mechanics~\cite{jaynes}. Thus, there arises the
interesting question of whether a given system of probability
distributions admits a representation in exponential form. Recent
work has shown that, in the case of finite-dimensional quantum
systems,  the time average of the density matrix can be expressed as
a grand canonical state, which assumes an exponential
form~\cite{Brody}. Motivated by this result, in the present paper we
derive a general theorem stating that an arbitrary system of
discrete or continuous probability densities admits a representation
in the form of an exponential family. This is surprising in that
even power-law distributions are thereby representable in
exponential form.

The paper is organised as follows. We first establish the result for
discrete and finite probability densities. An example of this result
has been demonstrated in \cite{Brody}; the purpose here is to
provide a simpler derivation of the general result. We then proceed
to consider the exponential representation for an arbitrary smooth
positive probability density function $\pi(x)$, and show that an
expression of the form $\pi(x)=\exp\left(-\sum_k\beta_k x^k\right)$
is always possible.

1. We begin our analysis in the case of a finite-dimensional
discrete probability distribution. Let $H$ be a random variable
assuming distinct values $\{E_i\}_{i= 0,1,\ldots n}$ with
probabilities $\{\pi_i\}_{i=0,1,\ldots n}$. Then, there is a
linearly independent family of $n$ random variables, including $H$
itself, such that any of these random variables can be expressed as
a function of $H$. There is a freedom in the choice of the family;
here, for simplicity, we choose the powers of $H$; thus, our family
of independent random variables is just the set
$\{1,H,H^2,H^3,\cdots,H^n\}$. The linear independence of these
random variables. i.e. the fact that the matrix of powers
$\{E_i^k\}$ is nonsnigular, follows from the elementary fact that an
n-th order polynomial vanishing at n+1 distinct points must be
identically zero. Moreover, the powers $H^m$ for all $m>n$ are
obviously expressible as linear combinations of powers
$\{H^k\}_{k=0,1,\ldots n}$. We define the moments
$\{\mu_k\}_{k=0,1,\ldots n}$ of $H$ by
\begin{eqnarray}
\mu_k = \sum_{i=0}^n \pi_i E_i^k, \label{eq:1}
\end{eqnarray}
where $\mu_0=1$. To establish the existence of an exponential
representation for $\{\pi_i\}$ two further ingredients are needed;
the first is the logarithmic entropy of Shannon and Wiener, defined
by
\begin{eqnarray}
S= -\sum_{i=0}^n \pi_i \ln \pi_i. \label{eq:2}
\end{eqnarray}
The second is the family of variables $\{\beta_k\}_{ k=1,\ldots n}$
conjugate to the moments $\{\mu_k\}$ with respect to the entropy $S$
in the sense that
\begin{eqnarray}
\beta_k = \frac{\partial S}{\partial \mu_k}. \label{eq:3}
\end{eqnarray}
We then have the following result:

\begin{prop}
The family of probabilities $\{\pi_i\}_{i=0,1,\ldots n}$ introduced
above can be expressed in the exponential form
\begin{eqnarray}
\pi_i = \exp\left(-\sum_{k=1}^n \beta_k E_i^k - \ln
Z({\boldsymbol{\beta}}) \right), \label{eq:4}
\end{eqnarray}
where $Z({\boldsymbol{\beta}})=\sum_{i=0}^n\exp\left(-\sum_{k=1}^n
\beta_k E_i^k\right)$.
\end{prop}

Since the matrix $\{E_i^k\}$  is nonsingular, equations (\ref{eq:1})
can be solved to express the $\{\pi_i\}$ as linear functions of the
moments $\{\mu_k\}$. That is, we can write
\begin{eqnarray}
\pi_i = \sum_{j=0}^n c_{ij} \mu_j, \label{eq:5}
\end{eqnarray}
where the constant coefficient matrix $\{c_{ij}\}$ is just the
inverse of the matrix $\{E_i^k\}$. Since the entropy is a concave
function of the moments $\{\mu_k\}$, the conjugate variables
$\{\beta_k\}$ introduced in (\ref{eq:3}) are in one-to-one
correspondence with $\{\mu_k\}$. In other words, (\ref{eq:3})
defines a Legendre transform~\cite{cencov}. Thus, in principle we
can express the moments $\{\mu_k\}$ in terms of the conjugate
variables $\{\beta_k\}$, substitute the results in (\ref{eq:5}), and
express the probabilities $\{\pi_i\}$ in terms of the variables
$\{\beta_k\}$. The proposition above states that the result of this
nonlinear transform can be expressed analytically, and is given by
an exponential family of distributions.

Proof. Since the row vectors $|i\rangle= (E_i^0,E_i^1,E_i^2,
\cdots,E_i^n)$ for $i=0,1,\ldots,n$ are linearly independent, we can
express the vector $-\ln\pi_i\in {\mathbb R}^{n+1}$ in the form
\begin{eqnarray}
-\ln\pi_i = \sum_{k=0}^n \beta_k E_i^k \label{eq:6}
\end{eqnarray}
for some coefficients $\{\beta_k\}$. Substituting (\ref{eq:6}) in
(\ref{eq:2}), we obtain
\begin{eqnarray}
S = \sum_{k=0}^n \beta_k \mu_k , \label{eq:7}
\end{eqnarray}
from which we deduce (\ref{eq:3}) \textit{a posteriori}. Finally,
solving (\ref{eq:6}) for $\pi_i$ we obtain the desired form
(\ref{eq:5}), where the normalisation condition for $\{\pi_i\}$
implies that $\beta_0 = \ln Z ({\boldsymbol{\beta}})$. \hfill$\Box$

By the above result, the nonlinear transform (\ref{eq:3}) can be
inverted analytically in the  form
\begin{eqnarray}
\mu_k = \sum_{i=0}^n E_i^k \exp\left(-\sum_{l=0}^n \beta_l E_i^l
\right). \label{eq:8}
\end{eqnarray}

2. An exponential representation can also be derived in the case of
an arbitrary smooth probability density function. In the continuous
case, however, the moments of the distribution need not exist in
general. Therefore, some of the preceding constructions involving
entropy and moments must be altered. We state the main result first:

\begin{prop}
Let $\pi(x)$ be a probability density function on the real line such
that $\ln\pi(x)$ is quadratically integrable with respect to the
Gaussian measure $\re^{-x^2} \rd x$. Then $\pi(x)$ can be expressed
in the exponential form
\begin{eqnarray}
\pi(x) = \exp\left(-\sum_{k=1}^n \beta_k x^k - \ln
Z({\boldsymbol{\beta}}) \right), \label{eq:9}
\end{eqnarray}
where $Z({\boldsymbol{\beta}})=\int_{-\infty}^\infty \exp\left( -
\sum_{k=1}^n \beta_k x^k\right)\rd x$, and where the value of $n$
may be infinite. The parameters $\{\beta_k\}$ are uniquely
determined by $\pi(x)$.
\end{prop}

The statement of Proposition~2 is perhaps surprising, because the
representation (\ref{eq:9}) applies, for example, to power-law
distributions such as the Cauchy distribution $1/[\pi(1+x^2)]$ for
which none of the moments exists. The proof goes as follows.

Proof. Since by assumption $\ln\pi(x)$ is quadratically integrable
with respect to the Gaussian measure, one can expand
$\ln\pi(x)\in{\cal L}^2({\mathbb R},\re^{-x^2} \rd x)$ in terms of
the Hermite polynomials $\{H_k(x)\}$, that is,
\begin{eqnarray}
\ln\pi(x) = - \sum_{k} \gamma_k H_k(x), \label{eq:10}
\end{eqnarray}
where
\begin{eqnarray}
\gamma_k = -\frac{1}{\sqrt{\pi}2^k k!} \int_{-\infty}^\infty
\ln\pi(x) H_k(x)\, \re^{-x^2} \rd x. \label{eq:11}
\end{eqnarray}
The infinite series in the right side of (\ref{eq:10}) converges
almost everywhere, since the squared Hilbert space norm
$\sum_k\gamma_k^2$ converges by assumption. Next, define a set of
numbers $\{\beta_k\}$ by the prescription
\begin{eqnarray}
\sum_k \beta_k x^k = \sum_k \gamma_k H_k(x). \label{eq:12}
\end{eqnarray}
Substituting this in (\ref{eq:10}) and solving the result for
$\pi(x)$, we deduce (\ref{eq:9}), where the normalisation condition
implies that $\beta_0 = \ln Z$. \hfill$\Box$

Can we establish a relation analogous to (\ref{eq:3}) for a general
probability density function? To this end we introduce what might
appropriately be called the `Gaussian moments' of $\pi(x)$ by
defining
\begin{eqnarray}
\mu_k = \int_{-\infty}^\infty x^k \pi(x)\, \re^{-x^2} \rd x.
\label{eq:13}
\end{eqnarray}
Similarly, we define the `Gaussian entropy' of $\pi(x)$ by
\begin{eqnarray}
S = -\int_{-\infty}^\infty \pi(x)\ln\pi(x)\, \re^{-x^2} \rd x.
\label{eq:14}
\end{eqnarray}
The coefficients $\{\beta_k\}$ appearing in (\ref{eq:9}) are then
related to the Gaussian moments $\{\mu_k\}$ defined by (\ref{eq:13})
via relation (\ref{eq:3}), provided we use the Gaussian entropy
(\ref{eq:14}).

The family of density functions for which $\ln\pi(x)$ is
quadratically integrable with respect to the Gaussian measure is
fairly large and includes, in particular, all the power-law
distributions. However, this family is not exhaustive. Nevertheless,
the representation (\ref{eq:9}) can be established for a much wider
class of density functions. The idea is to extend the formulation
based on the Gaussian measure into the class $\mathfrak{S}$ of
positive Schwartz functions (by this we mean functions that have
infinite numbers of derivatives, each of which decays faster than
any inverse polynomial). This class forms a convex cone which
includes, in particular, the Gaussian function $\re^{-x^2}$. Let
$s(x)\in \mathfrak{S}$ be a positive Schwartz function such that
$s(x)\ln \pi(x)$ is quadratically integrable with respect to the
Lebesgue measure. We then construct orthonormal polynomials
$\{J_k(x)\}$ in ${\mathcal L}^2({\mathbb R},s^2(x)\rd x)$ by means
of the Gram-Schmidt process. Approximating by integration over a
finite interval, we can then apply the Weierstrass approximation
theorem to establish the completeness of the set $\{J_k(x)\}$. The
function $\ln\pi(x)$ can therefore be expanded in a form analogous
to (\ref{eq:10}), with almost everywhere convergence. The
coefficients $\{\gamma_k\}$ depend upon the choice of the Schwartz
function $s(x)$, whereas the expansion coefficients $\{\beta_k\}$
defined in a manner analogous to (\ref{eq:12}) are basis
independent.

To show that the representation (\ref{eq:9}) is valid for all smooth
density functions we proceed as follows. First, we observe that
since $\pi(x)$ is nonnegative, $[1+(\ln\pi(x))^2]^{-1}$ is less than
or equal to one for all $x$. Therefore, the function
$f(x)=s(x)/[1+(\ln\pi(x))^2]$, for any $s(x)\in \mathfrak{S}$,
decays faster than any inverse polynomial. Thus, for an arbitrary
smooth density function $\pi(x)$, the logarithm $\ln\pi(x)$ is by
construction quadratically integrable in ${\mathcal L}^2({\mathbb R}
,f(x)\rd x)$. Of course, the density function could be so perverse
that $f(x)$ does not belong to $\mathfrak{S}$, i.e. the derivatives
of $f(x)$ need not decay faster than any inverse polynomial.
However, the behaviour of these derivatives is immaterial for our
construction, since we merely require that all polynomials are
quadratically integrable with respect to the measure $f(x)\rd x$.
Consequently, the above exponential representation is indeed valid
for all smooth density functions.

In statistics, the exponential family of distributions is generally
defined as the totality of density functions that admit
representations of the form $\exp(-\sum_{k=0}^n \beta_k T_k(x))$ for
a set of functions (sufficient statistics) $\{T_k(x)\}$, where $n$
is usually assumed finite. The foregoing result thus implies that
\textit{the exponential family of distributions is dense in the
totality of probability distributions}. Thus, the study of
probability distributions could, in principle, be restricted to the
exponential type. In specific applications, the practicality of this
depends upon the density function $\pi(x)$ and the choice of the
Schwartz function $s(x)$, since the rate of convergence depends upon
these ingredients.

From the physical point of view, the result established here also
leads to an interesting observation concerning equilibrium
properties of generic dynamical systems. We note that if a dynamical
system is in equilibrium, then the associated equilibrium
distribution is necessarily an energy distribution, since steady
state solutions to the Liouville equation (or the Heisenberg
equation in the case of a quantum system) are given by functions of
the Hamiltonian. Thus, we conclude that \textit{if a dynamical
system is in equilibrium, then the relevant equilibrium distribution
is necessarily expressible in grand canonical form}. We emphasise
that this result applies not only to thermal equilibrium, but to any
form of equilibrium state of a dynamical system.

\vspace{0.5cm}
\begin{footnotesize}
\noindent DCB acknowledges support from The Royal Society.
\end{footnotesize}
\vspace{0.5cm}

%\vskip8.0pt

\end{document}